\documentclass[pra,twocolumn,superscriptaddress,showpacs,floatfix]{revtex4-1}
\usepackage{epsfig,graphicx,amsmath,amsfonts,amssymb,times}
\usepackage[sort&compress]{natbib}
\begin{document}
	
\title{Multipartite Entanglement Generation Assisted by Inhomogeneous Coupling }
\date{\today}

\begin{abstract}
We show that controllable inhomogeneous coupling between two-level systems and a common data bus provides a fast mechanism to produce multipartite entanglement. Our proposal combines resonant interactions and engineering of coupling strengths---between the qubits and the single mode---leading to well defined entangled states. Furthermore, we show that, if the two-level systems interact dispersively with the quantized mode, engineering of coupling strengths allows the controlled access of the symmetric Hilbert space of qubits.
\end{abstract}

\author{C. E. L\'opez}
\affiliation{Departamento de F\'{\i}sica, Universidad de Santiago de
Chile, USACH, Casilla 307 Correo 2 Santiago, Chile}
\affiliation{Center for the Development of Nanoscience and Nanotechnology, 9170124, Estaci\'on Central, Santiago, Chile}

\author{F. Lastra}
\affiliation{Departamento de F\'{\i}sica, Universidad de Santiago de
Chile, USACH, Casilla 307 Correo 2 Santiago, Chile}

\author{G. Romero}
\affiliation{Departamento de Qu\'{\i}mica F\'{\i}sica, Universidad del Pa\'{\i}s Vasco-Euskal Herriko Unibertsitatea, Apartado 644, 48080 Bilbao, Spain}

\author{E. Solano}	
\affiliation{Departamento de Qu\'{\i}mica F\'{\i}sica, Universidad del Pa\'{\i}s Vasco-Euskal Herriko Unibertsitatea, Apartado 644, 48080 Bilbao, Spain}
\affiliation{IKERBASQUE, Basque Foundation for Science, Alameda Urquijo 36, 48011 Bilbao, Spain}

\author{J. C. Retamal}
\affiliation{Departamento de F\'{\i}sica, Universidad de Santiago de
Chile, USACH, Casilla 307 Correo 2 Santiago, Chile}
\affiliation{Center for the Development of Nanoscience and Nanotechnology, 9170124, Estaci\'on Central, Santiago, Chile}
\maketitle

{\it Introduction.}---Nowadays, complex quantum architectures are being developed aiming at physical implementations in quantum information processing~\cite{QC,hanson06,blais04,wallraff04,Chiorescu04,koch07,dicarlo10,abanto,kubo10,schuster10}. Their complexity is often revealed by intrinsic inhomogeneities which may limit the quantum control, for example, to prepare entangled states. This situation is found in a quantum dot, where the spatial distribution of the single-electron wave function makes its coupling to the spin bath position-dependent~\cite{hanson06}. Inhomogeneities also appear in circuit QED~\cite{blais04, wallraff04,Chiorescu04}, where the coupling strength of light-matter interaction depends on intrinsic parameters of qubits~\cite{koch07,dicarlo10}. Other physical scenarios exhibiting inhomogeneities are impurities embedded in silicon~\cite{abanto}, as well as hybrid systems involving NV-centers~\cite{kubo10,schuster10}. Nonetheless, the advent of these platforms---particularly circuit QED---brings new possibilities for engineering direct or indirect coupling among qubits in order to produce multipartite entangled states~\cite{2qbit3qbit}. Most implementations are based on slow off-resonant interaction among the qubits and a common data bus, however, resonant interactions have the advantage that their gating time scales with the inverse of the coupling strength, resulting in faster two-qubit gates~\cite{Resonant}.

In this work, we show that engineering the inhomogeneities of the coupling strengths between the two-level systems and a single-mode resonator, provides a fast mechanism for generating multipartite entangled states. First, we discuss the simplest case of a two two-level system interacting inhomogeneously with a single-mode resonator, looking for the specific conditions to generate well defined amounts of entanglement. We extend this study to the multipartite case, where our proposal allows the generation of $W$ entangled states. Second, we consider a dispersive interaction among qubits and the single-mode resonator such that we the engineering of coupling strengths allows the controlled access of the symmetric space of qubits. In particular, we profit from the resulting quantum dynamics to propose a mechanism to produce different classes of entangled states. Finally, we present our concluding remarks.

{\it Two-qubit system interacting inhomogeneously with a single quantized mode.}---Let us start our discussion considering the simplest case of a two-qubit system interacting with a single quantized mode. A main assumption of our model will be the individual addressing of qubits, allowing the controllability of their coupling strengths with the single mode. We aim to generate correlated states between qubits, so the physical mechanism has to be the exchange of a single excitation mediated by the quantum mode. If the first qubit is in the excited state, the second qubit in the ground state, and the quantized mode having no excitations, the single excitation located in the first qubit will be transferred among the parties according to the Hamiltonian	
\begin{equation}
\hat{H}=\hbar \chi_{1}(|e_{1}\rangle \langle g_{1}|\hat{a}+|g_{1}\rangle \langle
e_{1}|\hat{a}^{\dagger })+\hbar \chi_{2}(|e_{2}\rangle \langle g_{2}|\hat{a}+|g_{2}\rangle \langle
e_{2}|\hat{a}^{\dagger }),
\label{hamil1}
\end{equation}
where each qubit is coupled inhomogeneously ($\chi_1$ and $\chi_2$) to the quantum mode. As the initial condition is $|\psi_0\rangle=|e_1g_2\rangle|0\rangle$, the accessible Hilbert space for the composed system is spanned by vectors $\{|e_{1}g_{2}\rangle |0\rangle ,|g_{1}g_{2}\rangle |1\rangle,|g_{1}e_{2}\rangle |0\rangle \}$, so that the state of the system reads
\begin{equation}\label{psi1}
|\psi (t)\rangle =c_{1}(t)|e_{1}g_{2}\rangle |0\rangle
+c_{2}(t)|g_{1}g_{2} \rangle |1\rangle +c_{3}(t)|g_{1}e_{2}\rangle |0\rangle.
\end{equation}

The quantum dynamics supported by Hamiltonian~(\ref{hamil1}) induces entanglement between the qubits through their interaction with the quantum bus. However, it would be noticeable to produce maximal entanglement between the qubits in a single step. From Eq.~(\ref{psi1}), a maximal entanglement could be achieved when the probability amplitude $c_2 (t)$ vanishes. The probability amplitudes in Eq.~(\ref{psi1}) are given by
\begin{eqnarray}
c_{1}(t)&=&1+\frac{\chi_{1}^2}{\mu ^2}[\cos\mu t -1 ] \\
c_{2}(t)&=&-i\frac{\chi_{1}}{\mu }\sin\mu t \\
c_{3}(t)&=&\frac{\chi_{1} \chi_2 }{\mu ^2}[\cos\mu t -1 ],
\end{eqnarray}
where $\mu =\sqrt{ \chi_1^2 +\chi_2^2 } $. Imposing the  condition $c_2 (t)=0$, we obtain $c_{1} (t)=1-2 \chi_{1}^2 / \mu ^2$ and  $c_{3}(t)=-2 \chi_{1} \chi_2 / \mu ^2$. Thus resulting in the entangled state
\begin{equation}\label{psi2}
|\psi (t= \pi/\mu) \rangle = (1-2\frac{\chi_1 ^2}{\mu ^2}) |e_1 g_2 \rangle-2 \frac{\chi_{1} \chi_2}{ \mu ^2 } |g_1 e_2 \rangle.
\end{equation}

We realize that a maximally entangled state cannot be achieved in a single step with resonant interactions, and both qubits having same the coupling strengths. Here the inhomogeneous coupling will play a key role. In order to reach a maximally entangled state $|\psi (0) \rangle =(|e_1 g_2 \rangle+|g_1 e_2 \rangle)/\sqrt 2 $, the coupling strengths, $\chi_1$ and $\chi_2$, have to satisfy the relation $\chi_2 =-(\sqrt 2 +1) \chi_1 $ or $\chi_2 =(\sqrt 2 -1) \chi_1 $. This simple analysis shows that a suitable control of couplings strengths can be a useful mechanism to control the access to the Hilbert space of a two-qubit system. The natural extension of this idea is the case of a $N$-qubit system.

Consider an array of $N$ qubits, interacting in-homogeneously with the quantum mode. At resonance, this situation can be described by the Hamiltonian
\begin{equation}
\hat{H}=\hbar \sum_{i}\chi_{i}(|e_{i}\rangle \langle
g_{i}|\hat{a}+|g_{i}\rangle \langle e_{i}|\hat{a}^{\dagger }).
\label{Hn}
\end{equation}
Let us sssume the $k$-th qubit initially in the excited state and the remaining qubits be in the ground state, that is, $|\psi (0) \rangle=|gg.. e_k .. gg \rangle$. For this initial condition, the evolution given by Hamiltonian (\ref{Hn}) leads to the state
\begin{eqnarray}
|\Psi (t)\rangle &=&c_{0}(t)|g_{1}g_{2}.. \rangle |1\rangle
+\sum_{i}c_{i}(t)|g_{1}g_{2} .. e_{i}g_{i+1} .. \rangle |0\rangle,\nonumber \\
\end{eqnarray}
where the probability amplitudes read
\begin{eqnarray}
{c}_{0} &=&-i \frac{\chi_k }{\mu}\sin \mu t, \nonumber \\
{c}_{j} &=&\delta_{jk}+\frac{\chi_j \chi_k }{\mu ^2}[\cos \mu t - 1],
\end{eqnarray}
with $\mu=\sqrt{\sum_j \chi_j^2}$, $c_k(0)=1$ and $c_{j\ne k}(0)=0$.
From these expressions we learn that inhomogeneous coupling allow us to achieve a maximally multipartite entangled $W$-state resonantly and in a single step,  by setting the interaction time to $t=\pi/\mu$ and $\delta_{jk}-2 \chi_j \chi_k / \mu^2 =1/\sqrt N $. This leads to the following condition for the coupling strengths
\begin{equation}\label{cond}
\frac{\sum_j \chi_j }{\sum_j \chi_j ^2}=(1-\sqrt N) \frac{1}{2\chi_k}.
\end{equation}

For example, if the first qubit is initially excited and has a coupling constant $\chi_1$, and the remaining qubits being in the ground state with a coupling constant $\chi$,  the  condition (\ref{cond}) is fulfilled when $\chi_1 =(1+\sqrt{N}) \chi$, resulting in the $W$ state
\begin{equation}
|\psi (t=\pi/\mu)\rangle=\frac{1}{\sqrt N}\sum_j |gg\dots e_j \dots gg \rangle |0\rangle.
\end{equation}

Thus, we have been able to find conditions under which in-homogeneity allow the preparation of maximally multipartite entanglement with a single resonant interaction. Nonetheless, this physical mechanism is limited to the generation of entangled states involving a single excitation. Naturally, other limitations are those imposed by dissipative effects acting on the qubit or the quantum mode.

\begin{figure}
\includegraphics[width=60mm]{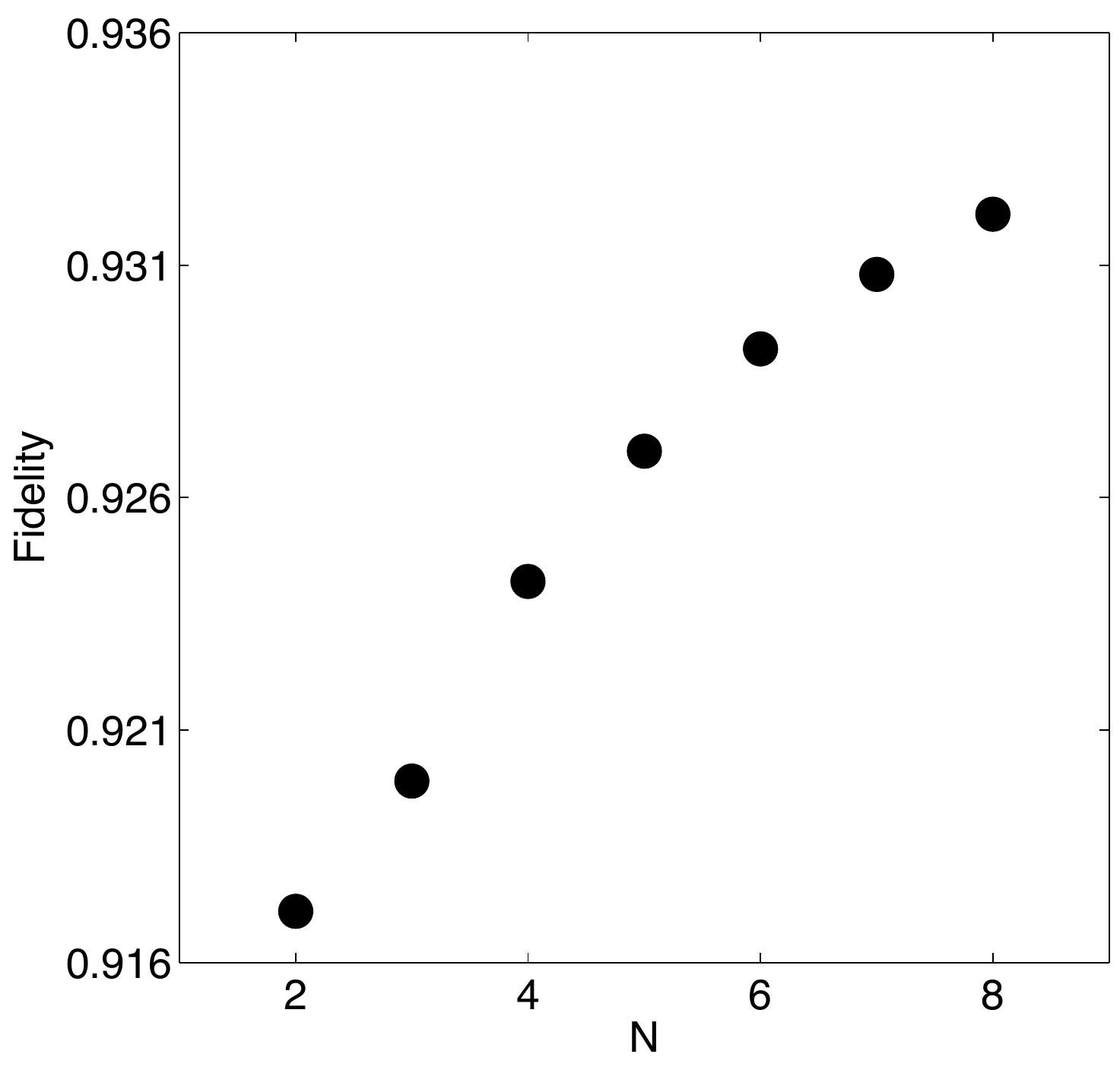}
\caption{Fidelity of maximally entanglement as a function of the number $N$ of qubits.  We have considered that the first qubit has the experimental parameters following Ref. \cite{wallraffPRL10}, while the remaining ones follow the corresponding engineered coupling values. Maximum coupling: $\chi_1 =54$ MHz, cavity $\kappa = 3.2$ MHz and the qubit spontaneous emission rate $\gamma=0.6$ MHz. The initial condition is $|\psi(0)\rangle=|gg...g\rangle|1\rangle$. }
\label{fig1}
\end{figure}

A physical setup where these ideas could be implemented, considering the available experimental resources, is circuit QED~\cite{blais04, wallraff04,Chiorescu04}. Two-qubit interactions mediated by a single mode of a resonator have been reported by several experimental groups~\cite{altomare10,wallraff09,schoelkopf10,majer07}. In particular, in Ref.~\cite{wallraff09}, the Tavis-Cummings model~\cite{tavis68} has been demonstrated experimentally up to three qubits, reaching maximally entangled states.

As an example, in Fig.~\ref{fig1}, the fidelity of the preparation of a $N$-particle $W$ state in a circuit QED setup is shown as a function of the number of qubits. In this simulation, we have considered dissipative mechanisms for cavity and qubits (dephasing and relaxation) using parameters of recent experiments~\cite{wallraffPRL10}. It is clear that the fidelity of the $W$ state increases as a function of the number of qubits, as expected from the higher effective coupling $\sim \sqrt{N} \chi$.

Another physical setup where maximal entanglement can be reached resonantly in a single step is trapped-ion systems~\cite{IonsReview}. Here, an alternative procedure can be followed: consider the first qubit tuned in the blue sideband transition and second qubit tuned in the red sideband one. For an initial condition $|\psi (0) \rangle=|g_1 g_2 \rangle$, the system will evolve within the subspace spanned by $\{|g_{1}g_{2}\rangle |0\rangle ,|e_{1}g_{2}\rangle |1\rangle,|e_{1}e_{2}\rangle |0\rangle\}$. Imposing similar conditions as for the previous case, this procedure will lead to the maximally entangled state $|\psi (t=\mu \pi)\rangle=(|e_1 e_2 \rangle + | g_1 g_2 \rangle )/\sqrt 2)$.

\begin{figure}
\includegraphics[width=60mm]{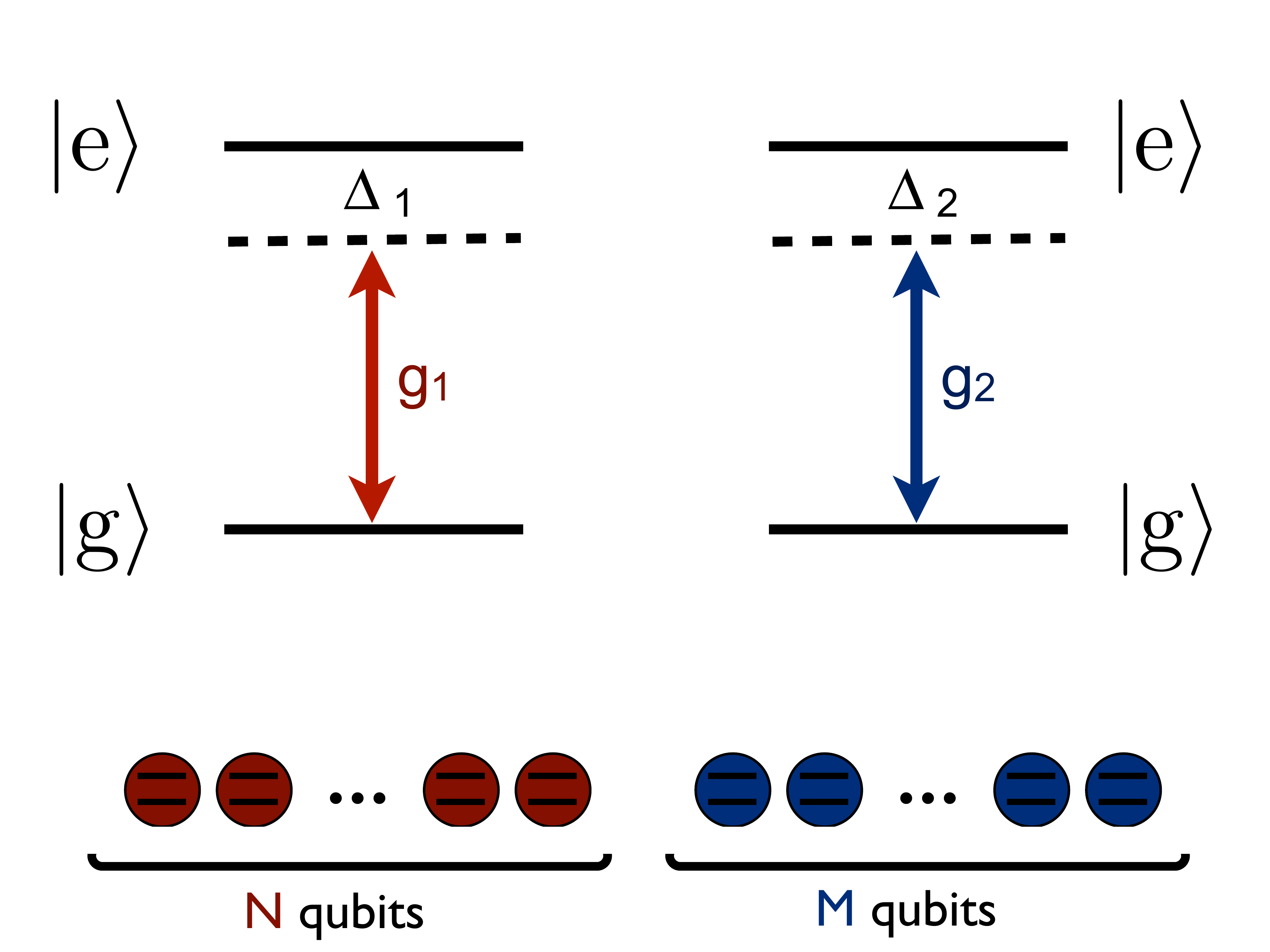}
\caption{(Color online) Schematic representation of $N$ qubits coupled to a single mode with coupling strength $g_{1}$ and other $M$ qubits with a coupling strength $g_{2}$.} \label{fig2}
\end{figure}

{\it Collective Control of Symmetric Dicke space.}---A resonant interaction between qubits and the quantum mode limits the possibility of generating other states than a $W$ state among qubits. As the quantum mode acts as an intermediary system, we can explore a regime where it can be
adiabatically eliminated, allowing a direct transfer of excitations between qubits. In this way, a set of qubits having a given number of $k$ excitations could directly interact with another set of
qubits having $q$ excitations, given rise to the generation of  states with $k+q$ excitations. To this purpose, we consider a collection of $N$ qubits, coupled to a quantum mode with coupling strengths $g_{1}$ and detuned from the qubit transition by $\Delta_{1}$. In addition, we consider $M$ qubits coupled to the same quantum mode with coupling strengths $g_{2}$ and detuning $\Delta_{2}$. This setup is schematically shown in Fig~\ref{fig2}. The Hamiltonian describing
this situation in the interaction picture reads
\begin{eqnarray}
\hat{H} &=& \hbar g_{1} \hat{a}\hat{S}^{\dag}e^{i \Delta_{1} t} + \hbar g_{2} \hat{a} \hat{J}^{\dag}e^{i \Delta_{2} t}+{\rm H.c.} ,
\end{eqnarray}
where we have defined the collective operators  $\hat{S}^{\dag} = \sum_{j=1}^{N} |{\rm e}\rangle_j \langle {\rm g}| $ and $\hat{J}^{\dag} = \sum_{j=N+1}^{M} |{\rm e} \rangle_j\langle {\rm g}|$.
In the far off-resonance regime, i.e. $\Delta_{1},\Delta_{2} \gg g_{1}, g_{2}$, an effective interaction between the $N$-qubit and the $M$-qubit subsystems arises. Under the assumption that  no photons are initially present in the quantum mode, the effective Hamiltonian reads

\begin{eqnarray}
\label{Heff1}
\hat{H}_{\rm eff}&=&\hbar \lambda_{1} \hat{S}^\dag \hat{S}+\hbar \lambda_{2} \hat{J}^\dag \hat{J} \notag \\
&+&\hbar \Omega_{\rm eff} \big[ \hat{S}^\dag \hat{J} e^{-i(\Delta_{1} -\Delta_{2})t}+\hat{S} \hat{J}^\dag e^{i(\Delta_{1} -\Delta_{2})t}\big],
\end{eqnarray}
where $\lambda_{j} = g_{j} ^2 / \Delta_{j} $, $\Omega_{\rm eff} = g_{1} g_{2} / \tilde{\Delta}$ and $\tilde{\Delta}=2 \Delta_1 \Delta_2 /(\Delta_1 + \Delta_2 )$. By noticing that
\begin{eqnarray}
\hat{S}^\dag \hat{S}&=& \sum_{i\neq j=1}^{N} \hat{\sigma}^{\dag}_i \hat{\sigma}_j+ \sum_{j=1}^N |e_j\rangle \langle e_j |, \\
\hat{J}^\dag \hat{J}&=& \sum_{i\neq j=N+1}^{M} \hat{\sigma}^{\dag}_i \hat{\sigma}_j+ \sum_{j=N+1}^M |e_j\rangle \langle e_j |,
\end{eqnarray}
the Hamiltonian (\ref{Heff1}) can be rewritten in a more convenient form $\hat{H}_{\rm eff}=\hat{H}_0+\hat{H}_I$, where
\begin{eqnarray}
\hat{H}_0 &=&\hbar \lambda_{1} \sum_{i\neq j=1}^{N} \hat{\sigma}^{\dag}_i \hat{\sigma}_j+\hbar \lambda_{2} \sum_{i\neq j=N+1}^{M} \hat{\sigma}^{\dag}_i \hat{\sigma}_j  \nonumber\\
\hat{H}_I&=&\hbar \Omega_{\rm eff} \left( \hat{S}^\dag \hat{J} e^{-i\delta t}+\hat{S} \hat{J}^\dag e^{i\delta t}\right),
\label{effecH}
\end{eqnarray}
where $\delta=\Delta_{2}-\Delta_{1} +\lambda_{2} - \lambda_{1}$ is the effective detuning parameter.  This Hamiltonian is the starting point for the subsequent discussion about the selective control of symmetric Dicke space. Indeed by conveniently choosing initial conditions for the set of $N$ and $M$ qubits we can force the evolution within symmetric Dicke space. In that follows, we will consider the general problem of the evolution of the set of $N$ and $M$ qubits, prepared in a special product of symmetric Dicke states. The same protocol will show us how to prepare a set of qubits in such states. Let us consider the overall system initially in a product state of the $N$-qubit subsystem in a symmetric state with $k$ excitations and the $M$-qubit subsystem having  $q$ excitations, that is
\begin{equation}\label{inidicke}
|\Psi_0 \rangle=|\mathrm{D}_{k} ^N \rangle |\mathrm{D}_{q} ^M \rangle ,
\end{equation}
where $|\mathrm{D}_{j} ^N \rangle$ stands for a $N$-particle Dicke state with $j$ excitations. In general, this initial state will evolve---under the action of the Hamiltonian (\ref{effecH})---within the subspace
\begin{equation}
\{|\mathrm{D}_{k} ^N \rangle |\mathrm{D}_{q} ^M \rangle,|\mathrm{D}_{k \pm 1} ^N \rangle |\mathrm{D}_{q \mp 1} ^M \rangle,|\mathrm{D}_{k \pm 2} ^N \rangle |\mathrm{D}_{q \mp 2} ^M \rangle,\dots\}.
\end{equation}

We aim to gain control of the symmetric space by reducing the size of the effective Hilbert space that the system can visit.  More precisely, given the initial state (\ref{inidicke}) we would like the evolution be restricted to the two-dimensional subspace: $\{|\mathrm{D}_{k} ^N \rangle |\mathrm{D}_{q} ^M \rangle,|\mathrm{D}_{k\pm1} ^N \rangle |\mathrm{D}_{q\mp1} ^M \rangle\}$. This can be done by adjusting the parameter $\delta$, setting this transition to resonance and leaving other possible transitions far from resonance. To see how this works let us analyze how the Hamiltonian (\ref{effecH}) acts on the initial state (\ref{inidicke})
\begin{eqnarray*}
\hat{H}_0 |\Psi_0 \rangle &=& \hbar\delta_{k,q} |\mathrm{D}_{k} ^N \rangle |\mathrm{D}_{q} ^M \rangle ,\\
\hat{H}_I |\Psi_0 \rangle &=& \hbar \Omega_{\rm eff} ( f_{k+1,q-1} e^{-i \delta t}|\mathrm{D}_{k+1} ^N \rangle |\mathrm{D}_{q-1} ^M \rangle \\
&&+  f_{k-1,q+1}e^{i \delta t} |\mathrm{D}_{k-1} ^N \rangle |\mathrm{D}_{q+1} ^M \rangle ),
\end{eqnarray*}
with
\begin{eqnarray}
f_{k,q}& \equiv& \sqrt{(k+1)(q+1)(N-k)(M-q)} ,\\
\delta_{k,q}& \equiv &\lambda_{1} k(N-k)+\lambda_{2} q(M-q).
\end{eqnarray}
Thus, each transition~$\{|\mathrm{D}_{k} ^N \rangle |\mathrm{D}_{q} ^M \rangle,|\mathrm{D}_{k\pm1} ^N \rangle |\mathrm{D}_{q\mp1} ^M \rangle\}$ has associated a detuning equal to $\widetilde{\delta}=\mp \delta-(\delta_{k \pm 1, q \mp 1}-\delta_{k,q})$.
Setting  $\widetilde{\delta} =0$ for given values of $k$, $q$, $N$ and $M$,  for a chosen transition other subspaces will be far from resonance, provided that this detuning is much larger than the effective coupling $\Omega_{\rm eff}$. As an example, let us consider the case when the first subsystem has single qubit in the excited state ($N=1$ and $k=1$) while the second subsystem is initially in a symmetric state with $M$ qubits and $q$ excitations, that is
\begin{equation}\label{ini1M}
|\Psi_0 \rangle = |e\rangle \sum_{j=0}^{M} a_j |\mathrm{D}_{j} ^M \rangle.
\end{equation}
Considering this initial state and setting the parameters such that $\widetilde{\delta}=\delta-\delta_{1,q+1}-\delta_{0,q}=0$, we enforce the system to evolves only in an effective two-dimensional Hilbert space spanned by $\{|e \rangle |\mathrm{D}_{q} ^M \rangle,|g \rangle |\mathrm{D}_{q+1} ^M \rangle\}$.  This selectivity is provided by $\widetilde{\delta} \gg \Omega_{\rm eff}$, such  that others subspaces do not evolve. When these requirements are fulfilled, the initial state (\ref{ini1M}) becomes
\begin{eqnarray}
|\Psi_t \rangle &=& |e\rangle \sum_{j\neq q} a_j |\mathrm{D}_{j} ^M  \rangle +a_q \big[ \cos{ (\Omega_{\rm eff} f_{0,q} t)}\notag \\
&+&  |e\rangle |\mathrm{D}_{q} ^M \rangle+i\sin{ (\Omega_{\rm eff} f_{0,q+1} t)} |g\rangle |\mathrm{D}_{q+1} ^M \rangle\big].\notag
\end{eqnarray}

\begin{figure}
\includegraphics[width=55mm]{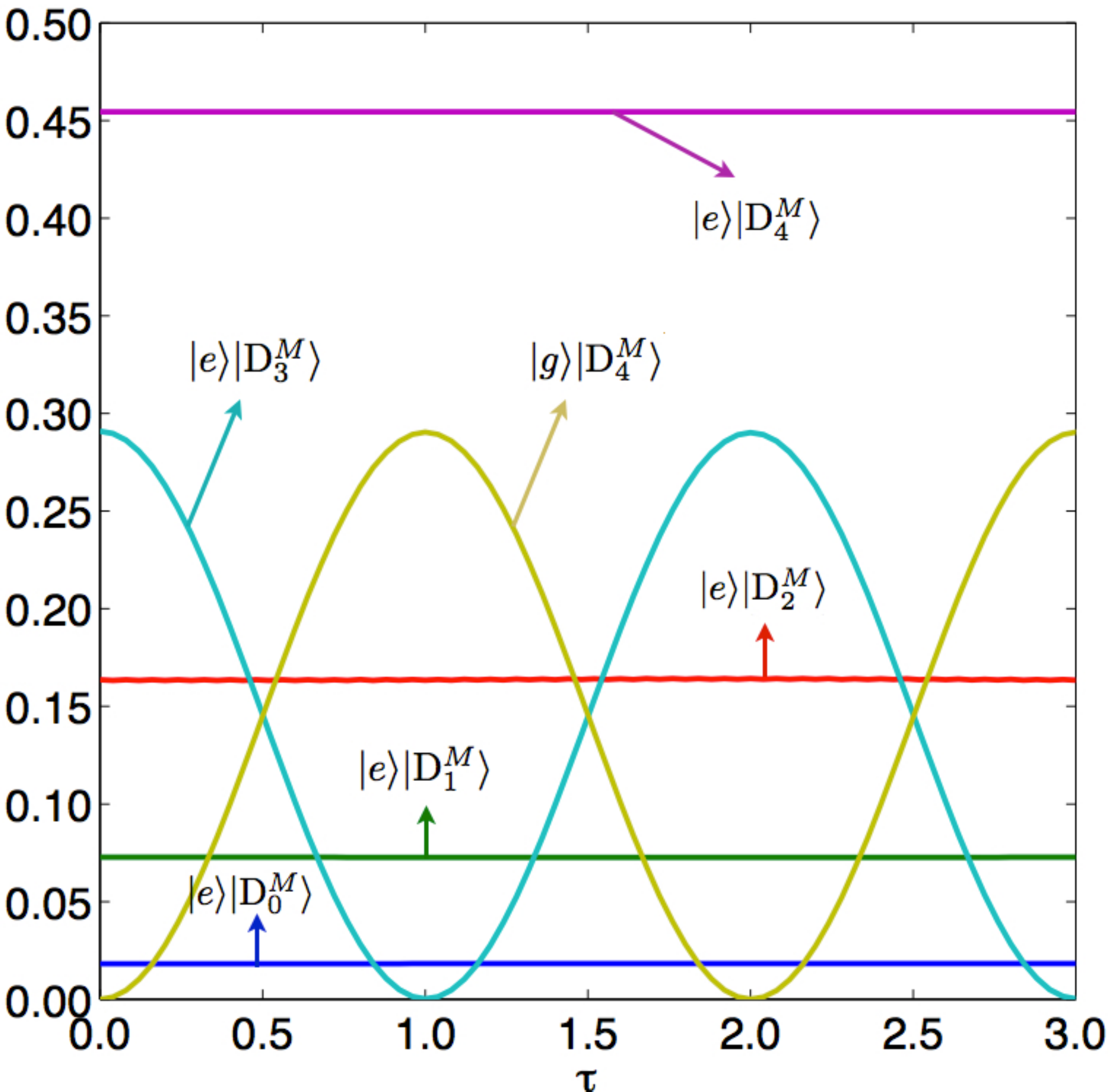}
\caption{(Color online) Evolution of populations of states composing the initial condition: $|\Psi_0 \rangle = (1/b)|e\rangle \sum_{j=0}^{M} a_j |\mathrm{D}_{j} ^M \rangle$ for M=6 as a function the time $\tau=2 f_{0,4} \Omega_{\rm eff} t/\pi$. The initial probability amplitudes are $a_0=0.1$, $a_1=0.2$, $a_2=0.3$, $a_3 = 0.4$, $a_4=0.5$ and $b$ is a normalization constant. The parameters are: $g_1 / g_2 =0.02$, $\Delta_2 /\Delta_1 =1.002$ and $\Delta_1 / g_2 =100$.} \label{fig3}
\end{figure}

For example, in Fig.~\ref{fig3} we show the dynamics for the case $M=6$ and $q=3$. Here we see that the dynamics occurs only in the selected subspace spanned by $\{|e \rangle |\mathrm{D}_{3} ^M \rangle,|g \rangle |\mathrm{D}_{4} ^M \rangle\}$, while the population of other states remaining virtually unchanged. This reduced dynamics can be used to prepare arbitrary Dicke states by measuring the state of the first qubit~\cite{lopez07}. That is, if we allow the system to evolve during a time $t=\pi/(2\Omega_{\rm eff} f_{0,q+1})$ and measuring the state of the first qubit we will find that the $M$-qubit subsystem will collapse into a Dicke state $|\mathrm{D}_{q+1} ^M \rangle$ with probability $|a_q|^2$.

However, preparation of arbitrary Dicke states can also be achieved deterministically using the ideas above. For example, we consider the system initially in the state
\begin{equation}\label{ini1M}
|\Psi_0 \rangle = |e\rangle  |\mathrm{D}_{q} ^M \rangle.
\end{equation}
Under the action of Hamiltonian (\ref{effecH}), this initial state will couple such as
\begin{eqnarray*}
\hat{H}_0 |\Psi_0 \rangle &=& \hbar\delta_{1,q} |e \rangle |\mathrm{D}_{q} ^M \rangle \\
\hat{H}_I |\Psi_0 \rangle &=& \hbar \Omega_{\rm eff}f_{0,q+1}  e^{i \delta t}|g \rangle |\mathrm{D}_{q+1} ^M \rangle,
\end{eqnarray*}
leading to an effective two-level dynamics in the subspace $\{|e \rangle |\mathrm{D}_{q} ^M \rangle,|g \rangle |\mathrm{D}_{q+1} ^M \rangle\}$. If we set the parameter such that, $\delta=\delta_{1,q}-\delta_{0,q+1}$, the evolution of the initial state will be
\begin{equation*}
|\Psi_t \rangle= \cos{(\Omega_{\rm eff} f_{0,q+1}t)}|e \rangle |\mathrm{D}_{q} ^M \rangle+i\sin{(\Omega_{\rm eff} f_{0,q+1}t)}|g \rangle |\mathrm{D}_{q+1} ^M \rangle.
\end{equation*}
When this state evolves during a time $t=\pi/(2\Omega_{\rm eff} f_{0,q+1})$, the resulting state will be a symmetric $M$-particle Dicke state with $q+1$ excitations
\begin{equation}
|\Psi_t \rangle =|g \rangle |\mathrm{D}_{q+1} ^M \rangle.
\end{equation}
Now, if the interaction time is set to be
\begin{equation}
\tau_{M,q}=\frac{1}{\Omega_{\rm eff} f_{0,q+1}}\arcsin{\sqrt{\frac{M-q}{M+1}}},
\end{equation}
the system will evolve into
\begin{equation}
|\Psi_t \rangle = |\mathrm{D}_{q+1} ^{M+1} \rangle
\end{equation}
that is, it becomes in a $(M+1)$-particle Dicke state with $q+1$ excitations.

Following this idea, arbitrary Dicke states can be deterministically prepared in a sequential manner. This works as follows: Consider and initial state of the form $|\Psi_0 \rangle = |e\rangle|\mathrm{D}_{0} ^M \rangle \equiv |e\rangle|g_1 g_2\dots g_{N} \rangle$. Then, if the system evolves during a time $t_1=\tau_{M,0}$, we will have $|\Psi_{t_1} \rangle = |\mathrm{D}_{1} ^{M+1} \rangle $. Now, we consider an additional qubit initially in the excited state, that is, $|\Psi_{t_1} \rangle \rightarrow  |e\rangle |\mathrm{D}_{1} ^{M+1} \rangle $. If this system evolves under the action of Hamiltonian (\ref{effecH}) during a time $t_2 = \tau_{M+1,1}$, the system will be in the state $|\Psi_{t_2} \rangle = |\mathrm{D}_{2} ^{M+2} \rangle $. In this manner, following this procedure we can sequentially prepare after $q$ steps, the Dicke state $|\mathrm{D}_{q} ^{M+q}\rangle$.

It is worth to mention that different proposals to prepare Dicke states have been presented for homogeneously coupled systems~\cite{lopez07,Linington08,Linington07,zheng1,solano1} and in inhomogeneous systems in a non-deterministic procedure~\cite{zheng2}.

{\it Entangled Dicke States.}---In the following we will show that this quantum dynamics can be used to prepare a different class of entangled Dicke states. Let us consider the situation described in
Fig.~\ref{fig2} for  $N	=M$. In such case we look for the conditions
required to achieve an effective two-level dynamics in a subspace
$\{|\mathrm{D}_{k} ^N \rangle |\mathrm{D}_{q} ^N
\rangle,|\mathrm{D}_{k\pm1} ^N \rangle |\mathrm{D}_{q\mp1} ^N
\rangle\}$. It is clear that under these conditions we should be
able to generate entanglement between Dicke states. A particular state we are able to prepare using the procedures described above is a NOON-like state of the form
\begin{equation}\label{entangled}
|\Psi\rangle = \frac{1}{\sqrt 2}(|\mathrm{D}_{N} ^N \rangle |\mathrm{D}_{0} ^N \rangle+|\mathrm{D}_{0} ^N \rangle |\mathrm{D}_{N} ^N \rangle).
\end{equation}

This can be done as follows. First, we consider the initial state $|\Psi_0 \rangle = |\mathrm{D}_{N} ^N \rangle |\mathrm{D}_{0} ^N \rangle$.
Then, we set into resonance the transition in subspace $\{|\mathrm{D}_{N} ^N \rangle |\mathrm{D}_{0} ^N \rangle,|\mathrm{D}_{N-1} ^N \rangle |\mathrm{D}_{1} ^N \rangle\}$. Setting the interaction time such as $t = \pi / (4 \Omega_{\rm eff} f_{N-1,1})$, the state will evolve to
\begin{equation}
|\Psi_1 \rangle = \frac{1}{\sqrt 2}(|\mathrm{D}_{N} ^N \rangle |\mathrm{D}_{0} ^N \rangle+i|\mathrm{D}_{N-1} ^N \rangle |\mathrm{D}_{1} ^N \rangle).
\end{equation}
By setting parameters such that the subspace $\{|\mathrm{D}_{N-1} ^N \rangle |\mathrm{D}_{1} ^N \rangle,|\mathrm{D}_{N-2} ^N \rangle |\mathrm{D}_{2} ^N \rangle\}$ is resonant,  the state $|\mathrm{D}_{N} ^N \rangle |\mathrm{D}_{0} ^N \rangle$ does not evolve since its associated transition is far from resonance. We lead the system to evolve during a time $t = \pi / (2 \Omega_{\rm eff} f_{N-2,2})$, obtaining
\begin{equation}
|\Psi_2 \rangle = \frac{1}{\sqrt 2}(|\mathrm{D}_{N} ^N \rangle |\mathrm{D}_{0} ^N \rangle+i^2 |\mathrm{D}_{N-2} ^N \rangle |\mathrm{D}_{2} ^N \rangle).
\end{equation}
Along the evolution we transfer the excitations from the first set of qubits to the second set of qubits in such a way
\begin{equation}
|\mathrm{D}_{N-2} ^N \rangle |\mathrm{D}_{2} ^N \rangle\rightarrow|\mathrm{D}_{N-3} ^N \rangle |\mathrm{D}_{3} ^N \rangle \rightarrow\dots \rightarrow |\mathrm{D}_{0} ^N \rangle |\mathrm{D}_{N} ^N. \rangle
\end{equation}
In this manner, after $N$ steps the state (\ref{entangled}) can be
prepared. As an example, we consider the case of each system having
$N=3$ qubits. The required values for the parameters to set into
resonance the needed transitions---to prepare an entangled state of
the form (\ref{entangled})--- are listed in Table I.

The scheme for finding the exact values to select a given transition is as
follows: first, as pointed out, we must fulfill the following
requirement, $\Delta_{1},\Delta_{2} \gg g_{1}, g_{2}$. To tune
the selected transition we must set the parameters $g_{1}$, $\Delta_{1},\Delta_{2},g_{2}$ or $g_{2}$ in such way that the relation $\widetilde{\delta}=0$ is satisfied. To achieve this, we may choose either $g_{1}$ as a function of
$\Delta_{1},\Delta_{2},g_{2}$ or $g_{2}$ as a function of
$\Delta_{1},\Delta_{2},g_{1}$. The choice of these parameter must be such that $\widetilde{\delta} \gg
\Omega_{\rm eff}$ for the other possible transitions, preventing in way, evolution in subspaces different from the chosen one. A direct consequence of such requirements is that the qubits cannot be homogeneously coupled, that is, $g_1 \ne g_2$.
\begin{table}
\begin{center}\begin{tabular}{|c|c|c|c|c|c|}\hline Step & $\delta$ & $\Delta_2 / \Delta_1$ & $g_1 / g_2$ & $\Delta_1 / g_1$ & $\Delta_2 / g_2$ \\\hline I & $3(\lambda_1 + \lambda_2)$ & 0.9996 & 18.630 & 53.7 & 1000 \\\hline II & 0 & 1.0025 & 70.621 & 20 & 1416 \\\hline III & $-3(\lambda_1 + \lambda_2)$ & 1.0075 & 17.611 & 20.15 & 355 \\\hline \end{tabular} \caption{Parameters for the preparation of the state (\ref{entangled}).}
\end{center}
\label{table}
\end{table}

\begin{figure}
\includegraphics[width=55mm]{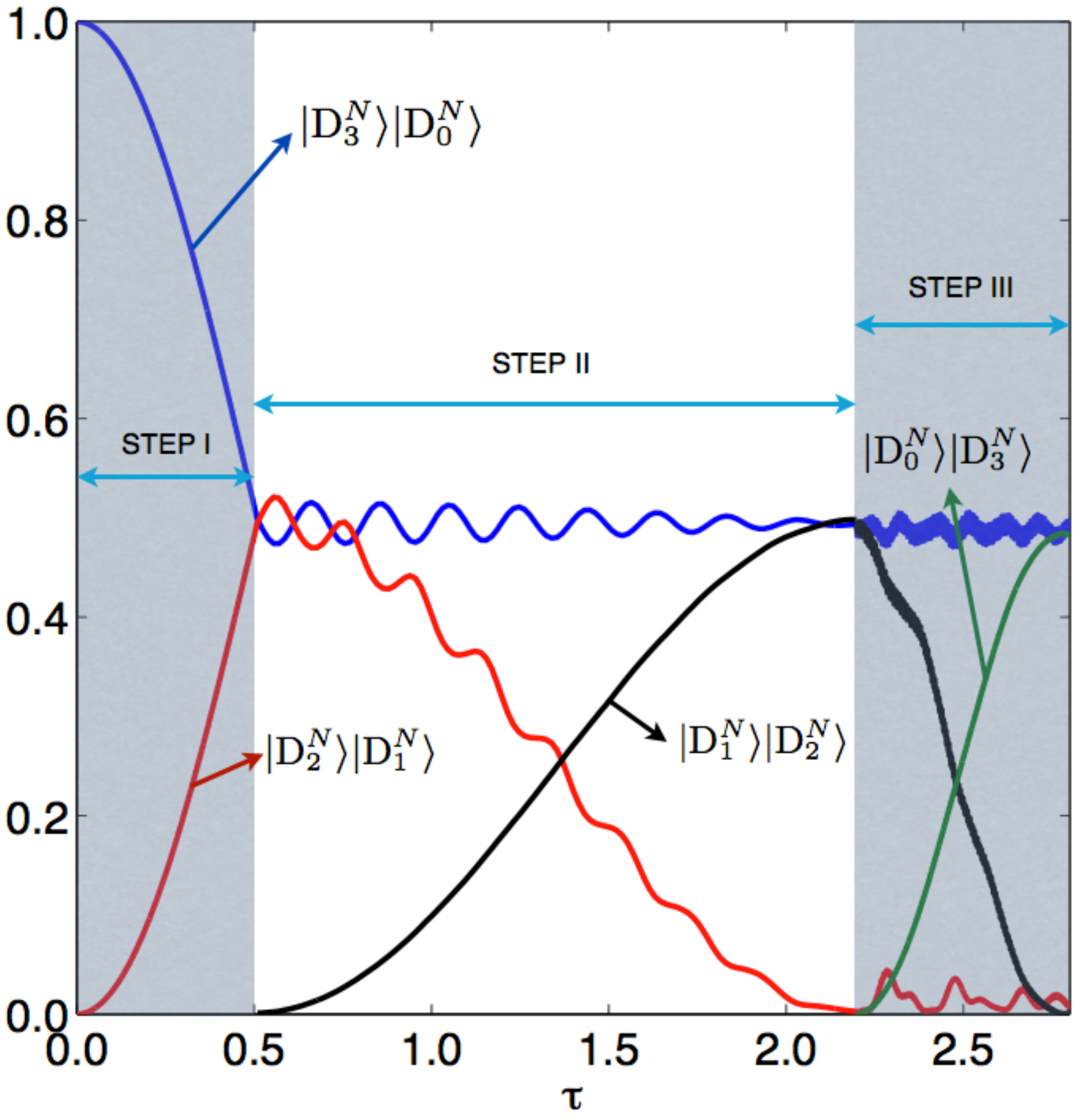}
\caption{(Color online ) Schematic of sequential preparation of NOON-like state in Eq.~(\ref{NOON-like}).} \label{fig4}
\end{figure}

In Fig.~\ref{fig4}, we show the process to prepare a multipartite entangled state $|\Psi\rangle = (1/\sqrt 2) (|\mathrm{D}_{3} ^N \rangle |\mathrm{D}_{0} ^N \rangle+|\mathrm{D}_{0} ^N \rangle |\mathrm{D}_{3} ^N \rangle)
$ for $N=3$. The first step---as shown in Fig.~\ref{fig4}---is to prepare a superposition
\begin{equation}
|\Psi\rangle = (1/\sqrt 2) (|\mathrm{D}_{3} ^N \rangle |\mathrm{D}_{0} ^N \rangle+|\mathrm{D}_{2} ^N \rangle |\mathrm{D}_{1} ^N \rangle).
\label{NOON-like}
\end{equation}
The following two steps allow the transitions
\begin{equation*}
|\mathrm{D}_{2} ^N \rangle |\mathrm{D}_{1} ^N \rangle\xrightarrow{\text{ STEP II}} |\mathrm{D}_{1} ^N \rangle |\mathrm{D}_{2} ^N \rangle\xrightarrow{\text{ STEP III}} |\mathrm{D}_{0} ^N \rangle |\mathrm{D}_{3} ^N \rangle,
\end{equation*}
without changing the population of the state $|\mathrm{D}_{3} ^N \rangle |\mathrm{D}_{0} ^N \rangle$.

{\it Conclusion.}---We have described the role of the inhomogeneous coupling between a single quantized mode and a set of two-level systems. In particular, we have shown that combining resonant interactions and engineering of coupling strengths allows a fast mechanism to produce multipartite entanglement. Furthermore, in the dispersive regime of light-matter interaction, the versatility in the engineering of coupling strengths provides a mechanism for controlling the access to the symmetric Hilbert space of qubits. At the same time, it allows the access to special classes of entangled states, both probabilistically and deterministically.

This work was supported by Financiamiento Basal para Centros Cient\'ificos y Tecnol\'ogicos de Excelencia, DICYT 041131LC, PBCT-CONICYT PSD54, Fondecyt 110700, Juan de la Cierva MICINN program, Basque Government Grant IT472-10, Spanish MICINN project FIS2009-12773-C02-01, SOLID and CCQED European projects.


\begin{thebibliography}{99}

\bibitem{QC} T. D. Ladd, F. Jelezko, R. Laflamme, Y. Nakamura, C. Monroe, J. L. O'Brien, Nature {\bf 464}, 45 (2010).

\bibitem{hanson06} R. Hanson, L. P. Kouwenhoven, J. R. Petta, S. Tarucha,and
L. M. K. Vandersypen, Rev. Mod. Phys. \textbf{79}, 1217 (2007).

\bibitem{blais04} A. Blais, R.-S. Huang, A. Wallraff, S. M. Girvin, and R. Schoelkopf, Phys. Rev. A {\bf 69}, 062320 (2004).

\bibitem{wallraff04} A. Wallraff {\it et al}, Nature (London) {\bf 431}, 162 (2004).

\bibitem{Chiorescu04} I. Chiorescu {\it et al.}, Nature (London) {\bf 431}, 159 (2004).

\bibitem{koch07} J. Koch {\it et al}, Phys. Rev. A {\bf 76}, 042319 (2007).

\bibitem{dicarlo10} L. DiCarlo {\it et al}, Nature {\bf 467}, 574 (2010).

\bibitem{abanto} M Abanto, L. Davidovich, Belita Koiller, and R. L. de Matos Filho, Phys. Rev. B \textbf{81} 085325 (2010).

\bibitem{kubo10} Y. Kubo {\it et al}, Phys. Rev. Lett. {\bf 105}, 140502 (2010).

\bibitem{schuster10} D. I. Schuster {\it et al}, Phys. Rev. Lett. {\bf 105}, 140501 (2010)

\bibitem{2qbit3qbit} J. H. Plantenberg {\it et al.}, Nature {\bf 447}, 836 (2007); J. Majer {\it et al.}, Nature {\bf 449}, 443 (2007); P. J. Leek, {\it et al.}, Phys. Rev. B {\bf 79}, 180511 (2009); L. Di Carlo {\it et al.}, Nature {\bf 460}, 240 (2009);R. C. Bialczak {\it et al.}, Nature Phys. {\bf 6}, 409 (2010); M. Neeley {\it et al.}, Nature {\bf 467}, 570 (2010).

\bibitem{Resonant} G. Haack, F. Helmer, M. Mariantoni, F. Marquardt, and E. Solano, Phys. Rev. B {\bf 82}, 024514 (2010).

\bibitem{altomare10} F. Altomare, J. I. Park, K. Cicak, M. A. Sillanp\"a\"a, M. S. Allman, D. Li, A. Sirois, J. A. Strong, J. D. Whittaker, R. W. Simmonds, Nature Phys. {\bf 6}, 777 (2010).

\bibitem{wallraff09} J. M. Fink, R. Bianchetti, M. Baur, M. G\"oppl, L. Steffen, S. Filipp, P. J. Leek, A. Blais, and A. Wallraff, Phys. Rev. Lett. {\bf 103}, 083601 (2009).

\bibitem{schoelkopf10} L. DiCarlo, M. D. Reed, L. Sun, B. R. Johnson, J. M. Chow, J. M. Gambetta, L. Frunzio, S. M. Girvin, M. H. Devoret, R. J. Schoelkopf, Nature {\bf 467}, 574 (2010).

\bibitem{majer07} J. Majer, J. M. Chow, J. M. Gambetta, Jens Koch, B. R. Johnson, J. A. Schreier, L. Frunzio, D. I. Schuster, A. A. Houck, A. Wallraff, {\it et al.}, Nature {\bf 449}, 443 (2007).

\bibitem{tavis68} M. Tavis and F. W. Cummings, Phys. Rev. \textbf{170}, 379
(1968).

\bibitem{wallraffPRL10} R. Bianchetti, S. Filipp, M. Baur, J. M. Fink, C. Lang, L. Steffen, M. Boissonneault, A. Blais, and A. Wallraff, Phys. Rev. Lett. {\bf 105}, 223601 (2010).

\bibitem{IonsReview} D. Leibfried, R. Blatt, C. Monroe, and D. Wineland, Rev. Mod. Phys. {\bf 75}, 281 (2003).

\bibitem{lopez07} C. E. L\'{o}pez, E. Solano and J. C. Retamal,
Phys. Rev. A \textbf{76}, 033413 (2007).

\bibitem{Linington08} I.E. Linigton, N.V. Vitanov, Phys. Rev. A \textbf{77}, 010302(R) (2008).

\bibitem{Linington07} I.E. Linigton, N.V. Vitanov, Phys. Rev. A \textbf{77}, 062327 (2007).

\bibitem{zheng1} Zhi-Biao Zheng Phys. Rev. A \textbf{77}, 033852 (2008).

\bibitem{solano1} C. Thiel, J. von Zanthier, T. Bastin, E. Solano, and G. S. Agarwal, Phys. Rev. Lett. {\bf 99}, 193602 (2007); T. Bastin, C. Thiel, J. von Zanthier, L. Lamata, E. Solano, and G. S. Agarwal, Phys. Rev. Lett. {\bf 102}, 053601 (2009). 

\bibitem{zheng2} Zhi-Biao Zheng J. Phys. B: At. Mol. Opt.Phys \textbf{42}, 085501(2009).

\end{thebibliography}
\end{document}